\documentclass[dvips,aas]{imsart}

\RequirePackage[OT1]{fontenc}
\RequirePackage[colorlinks]{hyperref}

\startlocaldefs
\newcommand{\beq}{\begin{equation}}
\newcommand{\eeq}{\end{equation}}
\newcommand{\cL}{{\mathcal{L}}}
\newcommand{\cW}{{\mathcal{W}}}

\def\Real{\hbox{I\kern-.1667em\hbox{R}}}
\def\Pr{\hbox{I\kern-.1667em\hbox{P}}}
\def\EE{\hbox{I\kern-.1667em\hbox{E}}}

\endlocaldefs

\begin{document}

\begin{frontmatter}
\title{Auditing a collection of races simultaneously\protect\thanksref{T1}}
\runtitle{Simultaneous Post-Election Audits}
\thankstext{T1}{I thank Mike Higgins, Mark Lindeman, Luke Miratrix and Ron Rivest for helpful conversations.}

\begin{aug}
\author{\fnms{Philip B.} \snm{Stark}\ead[label=e1]{stark@stat.berkeley.edu}}
\runauthor{P.B. Stark}

\affiliation{University of California, Berkeley}

\address{Department of Statistics, Code 3860\\
	Berkeley, CA 94720-3860\\
	\printead{e1}
}
\end{aug}

\begin{abstract}
        A collection of races in a single election can be audited as a group by auditing a random 
        sample of batches of ballots and combining observed discrepancies in the races 
        represented in those batches in a particular way:
        the maximum 
        across-race relative overstatement of pairwise margins (MARROP).
        A risk-limiting audit for the entire collection of races can be built on this 
        ballot-based auditing using a variety of probability sampling schemes.
        The audit controls the familywise error rate (the chance that one or more incorrect outcomes
        fails to be corrected by a full hand count) at a cost that can be lower than that of controlling
        the per-comparison error rate with independent audits.
        The approach is particularly efficient if batches are drawn with probability
        proportional to a bound on the MARROP (PPEB sampling).
\end{abstract}

\end{frontmatter}


\section{Introduction}
Post election audits can control the risk of certifying an election outcome that
disagrees with the outcome that a full hand count would show.
Pilot studies in California have shown that risk-limiting audits of individual races of a variety of
sizes can be conducted economically, within the canvass period~\cite{miratrixStark09a,ginnoldEtal09a}.
However, it is not efficient to audit a large number of races in a single election 
by simply repeating the audit process for each of those races.
The difficulty of auditing a large collection of races
is a logistical barrier to wider use of post-election audits to control risk.

This paper presents an approach to auditing an arbitrarily large number of
races in an election by hand-counting those races that appear on the ballots in a random
sample of batches of ballots.
Such ballot-based auditing is built into some state audit laws, such as California's ``1\% audit.''

In the new approach, for each batch of ballots in the sample, the discrepancies in the votes in the 
races represented in that batch are combined into a summary statistic, the maximum 
across-race relative overstatement of pairwise margins (MARROP).
This is a simple extension of the approach in~\cite{stark08d} to cover more than one race.
Any error that increased the apparent margin between some winner and some loser in a given race
is normalized by the apparent margin between those candidates.
The largest normalized error in a batch---maximized first across pairs of winners and losers for a given race
and then across races---summarizes 
the error in the batch.
This maximum across-race relative overstatement of pairwise margins can then be used with existing methods
designed for auditing individual races to limit the risk of certifying an incorrect outcome to $\alpha$, 
for instance, the methods introduced 
in~\cite{stark08a,stark09a,stark09b,miratrixStark09a}.
The result is a simultaneous risk-limiting audit of all the races: the audit limits the chance
that one or more incorrect outcomes will go uncorrected to at most $\alpha$.

This paper introduces the MARROP and
gives a cartoon application to a set of three races in an election in a jurisdiction roughly 
the size of a county.
The application uses the Kaplan-Markov bound~\cite{stark09b} for a sample drawn with probability 
proportional to an error bound~\cite{aslamEtal08,miratrixStark09a} to guarantee a known
minimum chance of a full hand count if that hand count would show that the outcome of any of
the races was wrong.
        
\section{Maximum Across-Race Relative Overstatement of Pairwise Margins (MARROP)}
As \cite{stark08d}~notes, for the apparent outcome of an election contest to be wrong, the
margin between some apparent winner of the contest and some apparent
loser of the contest must be overstated by at least 100\% of the margin between them.
Scaling errors by the margins they affect makes them commensurable.
This idea extends to several races:
for the apparent outcome of any of those races to be wrong, for some race,
the margin between some winner in that race and some loser in that race must be overstated 
by at least 100\%.

Suppose there are $N$ batches of ballots that together cover $R$ races.
Not every race is represented on every ballot, but together the $N$ batches include every ballot for
all $R$ races.
Race $r$ has $K_r$ ``candidates,'' which could be politicians or positions on an issue.
For instance, the ``candidates'' for a ballot measure might be ``yes on Measure~A'' and 
``no on measure~A.''
The total number of candidates or positions in all races is $K = \sum_{r=1}^R K_r$.
We take those $K$ candidates to be enumerated in some canonical order, for instance, alphabetically.

Voters eligible to vote in race $r$ may vote for up to $f_r$ candidates in that race
(race $r$ can have up to $f_r$ winners).
The $f_r$ candidates who apparently won race $r$ are those in $\cW_r$.
Those who apparently lost race $r$ are in $\cL_r$.
The apparent vote for candidate $k$ in batch $p$ is $v_{kp}$.
(If ballots in batch $p$ do not include the race $r$ in which candidate $k$ is competing, $v_{kp} \equiv 0$.)
The apparent vote for candidate $k$ is $V_k \equiv \sum_{p=1}^N v_{kp}$.
If candidates $w$ and $\ell$ are contestants in the same race $r$, the reported 
margin of apparent winner $w \in \cW_r$ over apparent loser $\ell \in \cL_r$ is 
\beq
   V_{w \ell} \equiv V_w - V_\ell > 0.
\eeq

The actual vote for candidate $k$ in batch $p$---the number of votes for $k$ that 
an audit would find---is $a_{kp}$.
If the ballots in batch $p$ do not include the race in which candidate 
$k$ is competing, $a_{kp} \equiv 0$.
The actual vote for candidate $k$ is $A_k \equiv \sum_{p=1}^N a_{kp}$.
If candidates $w$ and $\ell$ are contestants in the same race $r$, the actual margin of 
candidate $w \in \cW_r$ over candidate $\ell \in \cL_r$ is 
\beq
   A_{w\ell} \equiv A_w - A_\ell.
\eeq
All the apparent winners of all $R$ races are the true winners of those
races if 
\beq
	\min_{r \in \{1, \ldots, R\}} \; \min_{w \in \cW_r, \ell \in \cL_r} A_{w\ell} > 0.
\eeq
If $w \in \cW_r$ and $\ell \in \cL_r$, define
\beq
	e_{pw\ell} \equiv \left \{
            \begin{array}{ll}
	           \frac{(v_{w p} - v_{\ell p}) - (a_{w p} - a_{\ell p})}{V_{w \ell}}, & 
                             \mbox{ if ballots in batch $p$ contain race $r$} \cr
                    0, & \mbox{ otherwise.}
            \end{array}
       \right .
\eeq
For the actual outcome of any of the $R$ races to differ from its apparent outcome,
there must exist $r \in \{1, \ldots, R\}$, $w \in \cW_r$ and
$\ell \in \cL_r$ for which $\sum_{p=1}^N e_{pw\ell} \ge 1$.
The {\em maximum across-race relative overstatement of pairwise margins in batch $p$\/} is
\beq
	e_p \equiv \max_{r \in \{1, \ldots, R\}} \max_{w \in \cW_r, \ell \in \cL_r} e_{pw\ell}.
\eeq
Now
\beq
	\max_{r \in \{1, \ldots, R\}} \max_{w \in \cW_r, \ell \in \cL_r} \sum_{p=1}^N e_{pw\ell} \le
	\sum_{p=1}^N \max_{r \in \{1, \ldots, R\}} \max_{w \in \cW_r, \ell \in \cL_r} e_{pw\ell} = \sum_{p=1}^N e_p
        \equiv E.
\eeq
The sum on the right, $E$, is the {\em maximum across-race relative overstatement of pairwise margins\/} (MARROP).
If $E < 1$, the apparent electoral outcome of each of the $R$ races is the same outcome that
a full hand count would show.

Think of the family of $R$ null hypotheses, {\em the outcome of race $r$ is incorrect\/}.
Then $E<1$ is a sufficient condition for the entire family of $R$ null hypotheses to be false.
If an audit gives strong statistical evidence that $E < 1$, we can safely conclude that the 
apparent outcomes of {\em all\/} $R$ races are correct.
If we test the hypothesis $E \ge 1$ at significance level $\alpha$, that gives a test of the family
of $R$ hypotheses with familywise error rate no larger than $\alpha$.

Suppose the number of valid ballots cast in batch $p$ for race $r$ is at most $b_{rp}$.\footnote{%
   If the batches are homogeneous with respect to ballot style, then $b_{1p} = b_{2p} = \ldots = b_{Rp}$.
   This is the case when batches of ballots correspond to precincts.
   In some jurisdictions, however, VBM ballots are counted in ``decks'' that bear no special 
   relationship to geography.
   Then, the values of $b_{rp}$ for a single batch $p$ can depend on the race $r$.
}
Clearly $a_{wp} \ge 0$ and $a_{\ell p} \le b_{rp}$, if $\ell$ is a candidate in race $r$.
Hence, $e_{pw\ell} \le (v_{w p} - v_{\ell p} + b_{rp})/V_{w \ell}$, and so
\beq
	e_p \le 
	          \max_{r \in \{1, \ldots, R\}}
                        \max_{w \in \cW_r, \ell \in \cL_r} \frac{v_{w p} - v_{\ell p} + 
	                  b_{rp}}{V_{w \ell}}
             \equiv u_p.
\eeq
The bound $u_p$ is a limit on the relative overstatement of {\em any\/} margin 
that can be concealed in batch $p$.
If $U \equiv \sum_p u_p < 1$, the outcome of the election must be correct so no audit is needed.

Otherwise, if the values of $u_p$ are generally small, error sufficient to cause the wrong candidate to
appear to win any of the races must be spread out across many batches, while if some of the values of $u_p$ are 
large, outcomes can be wrong even if most batches show no error at all.
The values of $u_p$ can be used to perform NEGEXP sampling or sampling 
with probability proportional to an error bound (PPEB)~\cite{aslamEtal08}.

The values of $e_p$ observed in a random sample (simple, stratified, NEGEXP or PPEB) can be 
used to calculate a $P$ value for the compound hypothesis that one or more of the apparent 
outcomes of the $R$ races differs from the outcome that a full hand count of all the ballots 
in that race would show.
See~\cite{stark09b} for details.
Those $P$-value calculations can be embedded in a sequential procedure for testing whether one or more of the
outcomes is wrong, using approaches like that described by~\cite{stark09a}.
The resulting test controls the {\em familywise error rate\/} for testing the collection of 
hypotheses that the outcome of each race is correct.
That is, the test keeps small the chance of incorrectly concluding that the outcomes are 
correct when any of the outcomes is wrong.

The {\em taint\/} of batch $p$ is
\beq
   \tau_p = \frac{e_p}{u_p} \le 1.
\eeq
For PPEB samples, it is convenient to work with taint $\tau_p$ rather than with error $e_p$, because
the expected value of the taint in a batch drawn by PPEB is $E/U$.
See~\cite{stark09b,miratrixStark09a}.

\section{Illustration}
This section presents a cartoon of an election with $R=3$ contests in a jurisdiction that has 200~precincts.
Each of the three races has only two contestants.
Race~A is jurisdiction-wide; the overall result is 50\% for the apparent winner, 45\% for the apparent loser,
and 5\% undervotes and invalid ballots.
Race~B involves half the precincts in the jurisdiction; the overall result for this race is
50\% for the apparent winner, 40\% for the apparent loser, and 10\% undervotes and invalid ballots.
Race~C involves 60 of the precincts in the jurisdiction, of which 30 overlap with the second race.
The overall result for this race is 50\% for the apparent winner, 35\% for the apparent loser, and
15\% undervotes and invalid ballots.

The auditable batches of ballots comprise ballots cast either in-precinct (IP) or by mail (VBM) for each of
the 200~precincts in the jurisdiction; thus there are $N=400$~auditable batches of ballots in all.
For the sake of illustration, we take the IP batches to contain 400 ballots each and the VBM batches to 
contain 200~ballots each, and we assume that, for each race, the margins are the same in all 400~batches.
A summary is given in table~\ref{tab:race_results}.

\begin{table}
\begin{tabular}{r|rrr|rrr|rr|rr}
     &           &         &         &        &       &        &\multicolumn{2}{c|}{IP batches} & \multicolumn{2}{c}{VBM batches} \cr
Race & precincts & batches & ballots & winner & loser & margin & winner & loser & winner & loser \cr
\hline
A    & 200 & 400 & 120,000 & 60,000 & 54,000 & 6,000 & 200 & 180 & 100 & 90  \cr
B    & 100 & 200 &  60,000 & 30,000 & 24,000 & 6,000 & 200 & 160 & 100 & 80  \cr
C    &  60 & 120 &  36,000 & 18,000 & 12,600 & 5,400 & 200 & 140 & 100 & 70
\end{tabular}
\caption{\protect \label{tab:race_results}
Hypothetical reported results for an election with three overlapping races.}
\begin{minipage}{4.5in}
Race~A spans the entire jurisdiction, 200~precincts.
Race~B includes 100 of the precincts in the jurisdiction.
Race~C includes 60 of the precincts in the jurisdiction; 30 of those are also in race~B.
Each precinct is divided into two batches of ballots: 400 ballots cast in-precinct (IP)
and 200 ballots cast by mail (VBM).
In addition to valid votes for the candidates, there are undervotes and invalid ballots.
\end{minipage}
\end{table}

There are eight situations to consider in calculating $u_p$: IP versus VBM batches where 
voters can vote only in race~A, in races~A and~B, in races~A and~C, and in all three races.
Consider the last of these for an IP batch ($b_{rp} = 400$).
\begin{eqnarray}
  u_p &=& \max \left \{ 
                  \frac{200 - 180 + 400}{6,000}, \frac{200 - 160 + 400}{6,000}, \frac{200 - 140 + 400}{5,400}
               \right \} \nonumber \\
      &=& \max \{ 0.0700, 0.0733, 0.0852 \} = 0.0852.
\end{eqnarray}
For a VBM batch in which voters were eligible to vote in all three races ($b_{rp} = 200$),
\begin{eqnarray}
  u_p &=& \max \left \{ 
                  \frac{100 - 90 + 200}{6,000}, \frac{100 - 80 + 200}{6,000}, \frac{100 - 70 + 200}{5,400}
               \right \} \nonumber \\
      &=& \max \{ 0.0350, 0.0367, 0.0426 \} = 0.0426.
\end{eqnarray}
Table~\ref{tab:up} lists the values of $u_p$ for all eight cases.
The total of all the error bounds for all $N=400$ batches is
\beq
    U = 70\times (0.0700 + 0.0350 + 0.0733 + 0.0367) + 2\times 30 \times (0.0852 + 0.0426) = 22.718.
\eeq

\begin{table}
\begin{tabular}{l|rr}
batch type          & batches & $u_p$ \cr
\hline
IP--Race~A only     &    70   & 0.0700 \cr
VBM--Race~A only    &    70   & 0.0350 \cr
\hline
IP--Races~A and~B   &    70   & 0.0733 \cr
VBM--Races~A and~B  &    70   & 0.0367 \cr
\hline
IP--Races~A and~C   &    30   & 0.0852 \cr
VBM--Races~A and~C  &    30   & 0.0426 \cr
\hline
IP--Races~A, B \& C &    30   & 0.0852 \cr
VBM--Races~A, B \& C &   30   & 0.0426 \cr
\end{tabular}
\caption{\protect \label{tab:up} Upper bounds on the MARROP in each batch
for the eight kinds of batches of ballots in a hypothetical race.
}
\end{table}

Suppose we want to design a PPEB-based audit that has at least a 75\% chance of requiring a full hand 
count if a full hand count would show a different outcome for any of the three races.
That controls the risk (that an incorrect result will not be corrected by a full hand count) 
to be at most $\alpha = 0.25$.
We can base such an audit on the Kaplan-Markov approach in~\cite{stark09b}.
We draw $n$ times with replacement from the 400~batches.
In each draw, the chance of selecting batch $p$ is $u_p/U$.
The draws are independent.

Let $T_j$ be the taint of the $j$th draw, that is, $T_j = \tau_p \equiv e_p/u_p$ for the 
batch $p$ that is selected in the $j$th draw.
Define
\beq
   P \equiv \min_{j=1}^n \prod_{i=1}^j \frac{1 - 1/U}{1-T_i}.
\eeq
Then we can stop the audit without a full hand count if $P < \alpha = 0.25$~\cite{stark09b}.

In particular, suppose we make $n=36$ PPEB draws, 5~of which show taint $\tau_p = 0.04$ and the rest of
which show $\tau_p = 0$.\footnote{%
    Taint of 0.04 corresponds to a different number of errors in different batches, depending on the value
    of $u_p$ in the batch and the margin that the error affects.
    In an IP batch of ballots that includes race~C, an error that overstates the margin in race~A or 
    race~B by 20~votes is a taint of just under 0.04, while in a VBM batch of ballots that includes
    only race~A, an error that overstates the margin in race~A by 8~votes is a taint of just under 0.04.
}
Then $P = 0.243$: we could stop the audit without a full hand count.
The risk that the outcome of any of the three races is wrong is at most 25\% (and plausibly far lower,
since this approach makes a number of very conservative choices).

Note that the expected number of distinct batches drawn in the $n=36$ draws is 
\beq
   \sum_{p=1}^{400} [1- (1-u_p/U)^{36}] = 34.3,
\eeq
about 8.6\% of the 400~auditable batches.
However, those batches would tend to be the larger (IP) batches.
Let $b_p$ denote the number of ballots in batch $p$ 
($b_p = 400$ for IP batches and $b_p = 200$ for VBM batches).
The expected number of ballots audited is
\beq
   \sum_{p=1}^{400} b_p [1- (1 - u_p/U)^{36}] = 11,387.3,
\eeq
about 9.5\% of the 120,000 ballots.
The expected number of votes audited, 20,617.68, can be calculated analogously:
substitute in place of $b_p$ the number of
voting possibilities in batch $p$ (from 200 for VBM batches that include only race~A
up to 1,200 for IP batches that include all three races).

In contrast, suppose we were auditing only race~A.  
Then the error bounds would be $u_p = 0.07$ for the 200~IP batches and $u_p = 0.035$ for the 200~VBM batches;
The total error bound would be $U_A = 21$, a bit smaller than the previous value, $U = 22.718$.
If the sample taints in $n=36$ draws were as before---five equal to $0.04$ and 
31 equal to 0---the value of $P$ would be
$0.212$.
This is a bit smaller than the value $0.243$ for auditing all three races, stronger 
evidence that the outcome of that single race was correct.

Conversely, if we had made only $n=33$~draws and had seen five taints equal to $0.04$ and 28 equal to zero, 
the value of $P$ would be $0.245$, and we would be able to confirm the outcome of that single race with risk no 
greater than $\alpha = 0.25$.
The workload would be somewhat lower, both because we would be counting only one race on each ballot and because the 
number of batches drawn would be lower.
The expected number of batches audited
would be 31.6 versus 34.3, and the expected number
of ballots audited would be 9,778 versus 11,387.
But we would only be testing the outcome of race~A.

Suppose we audited all three races independently.
We have a choice to make about multiplicity---the fact that we are testing more than one hypothesis.
The simultaneous audit procedure based on MARROP
has the property that there is at least 75\% chance of a 
full hand count of every race that has an incorrect outcome, i.e., risk at most $\alpha = 0.25$ that 
one or more incorrect outcomes will be certified.
Suppose we choose to maintain this property---keeping the familywise error rate (FWER) at most $\alpha = 0.25$.
We split the risk across the three audits by requiring each to have chance at least $0.75^{1/3} = 0.909$ of
a full count if the outcome is incorrect. 
The chance all three will progress to full counts if all three outcomes are incorrect is then 
at least $0.909^3 = 0.75$.

We could instead control the per-comparison error rate (PCER) to be at most $\alpha = 0.25$.
That would mean that for each audited race, the chance of a full hand count if the outcome is wrong
is at least 75\%.
However, the chance that one or more of the three races escapes a full hand count can be greater than $0.25$
when more than one outcome is wrong.
This way of dealing with multiplicity is a bit unfair to MARROP, because MARROP in fact has a lower error rate.
Keeping the PCER below $0.25$ requires rather smaller sample sizes than keeping the FWER below $0.25$.

Table~\ref{tab:independent} lays out the total error bounds for auditing the three races separately and
the sample sizes that would be needed to stop the audits without a full count if the corresponding samples 
had at most five taints no larger than 0.04 and the rest of the taints were zero, while keeping the
familywise error rate (FWER) or the per-comparison error rate (PCER) under $\alpha = 0.25$.

\begin{table}
\begin{tabular}{r|r|rrrr|rrrr}
        &         & \multicolumn{4}{c}{FWER} & \multicolumn{4}{c}{PCER} \cr
        &         &       & expected & expected & expected   &     & expected & expected  & expected\cr
Race    &  $U$    &   $n$ & batches  & ballots  & votes      & $n$ & batches  & ballots   &  votes\cr
\hline
A       &   21.00 &    52 & 48.49    & 16,074.23 & 16,074.23 & 33  &  31.58   & 10,488.77 & 10,488.77\cr
B       &   11.00 &    28 & 26.01    &  8,615.69 &  8,615.69 & 17  &  16.27   &  5,402.16 &  5,402.16\cr
C       &    7.67 &    19 & 17.50    &  5,795.81 &  5,795.81 & 12  &  11.41   &  3,787.51 &  3,787.51\cr
\hline
all     &         &       & 85.13    & 28,038.26 & 30,485.73 &     &  56.38   & 18,649.98 & 19,678.44\cr
\hline
MARROP  &   22.72 &    36 & 34.30    & 11,387.29 & 20,617.68 &     &          &           &
\end{tabular}
\caption{\protect \label{tab:independent}Comparison of independent and simultaneous audits controlling FWER and PCER.}
\begin{minipage}{4.5in}
The familywise error rate (FWER) of a collection of audits is the chance that one or more fails
to result in a hand count when the corresponding outcome is incorrect.
If the FWER is at most $0.25$, the chance that there is a full hand
count of every race with an incorrect outcome is at least 75\%.
The per-comparison error rate of a collection of audits is the chance that each audit fails
to result in a hand count when the outcome of the race under audit is incorrect.
If the PCER is at most $0.25$, then, for each race, if the outcome is wrong, there is at 
least a 75\% chance of a full hand count.
However, the chance that there is a full hand count of every race with an incorrect outcome could be
less than 75\%: PCER is less stringent than FWER.
The total bounds on the error are given in column~2.
Suppose we design the audits to stop if no more than five nonzero taints of no more than 0.04 are 
observed; otherwise, the audit progresses to a full hand count.
To control the FWER, the number of draws is in column~3; the
expected number of distinct batches audited in column~4; the expected number of distinct
ballots audited in column~5; and the expected number of votes audited is in column~6.
To control the PCER, the number of draws is in column~7; the
expected number of distinct batches audited in column~8; the expected number of distinct
ballots audited in column~9; and the expected number of votes audited is in column~10.
The row labeled ``all'' gives the overall expected number of distinct batches and ballots
audited in the three independent audits to control the FWER or the PCER.
The row labeled ``MARROP'' gives the values for a simultaneous audit of all three races using
the maximum across race relative overstatement of pairwise margins, which controls the FWER to
be $0.25$ or below.
Far less work is required than using independent audits
the risk to the same level, measured by expected ballots or batches.
The expected number of votes is far less than required to control the FWER using independent audits, and
only slightly higher than required to control the PCER---even though the MARROP audit controls FWER.
\end{minipage}
\end{table}

How much work should we expect to do to audit all three races separately?
Let $u_{Ap}$ denote the error bound for batch $p$ if only race~A is audited.
Let $U_A = \sum_{p=1}^N u_{Ap}$.
Define $u_{Bp}$, $U_B$, $u_{Cp}$ and $U_C$ analogously.
The expected number of distinct batches that would be audited in all is
\beq
   \sum_{p=1}^N [1 - (1-u_{Ap}/U_A)^{n_A}(1-u_{Bp}/U_B)^{n_B}(1-u_{Cp}/U_C)^{n_C}],
\eeq
and the expected number of distinct ballots audited would be
\beq
   \sum_{p=1}^N b_p[1 - (1-u_{Ap}/U_A)^{n_A}(1-u_{Bp}/U_B)^{n_B}(1-u_{Cp}/U_C)^{n_C}].
\eeq
For some of those batches of ballots, only one race would be audited; for some, two races; 
and for some, all three.
See table~\ref{tab:independent} for numerical comparisons of MARROP against independent audits
that control FWER or PCER.
MARROP is much more efficient in this example.

The simultaneous approach based on the MARROP 
controls the overall risk with far less auditing effort.
The effort depends strongly on the number of batches as well as the number of votes, because there
are substantial logistical costs associated with pulling batches of ballots together for counting, and
there are economies in counting all the races on a single ballot.
Even though MARROP controls FWER, the workload is lower than for independent audits that only control 
PCER---a less stringent criterion---if work is measured by the number of batches or ballots audited.
(The number of votes audited is a bit higher than for independent audits that control PCER, but far lower
than for independent audits that control FWER.)

\section{Summary}
A collection of races can be audited simultaneously using the maximum
across-race relative overstatement of pairwise margins (MARROP).
Drawing batches using probability proportional to an upper bound on the MARROP---a form of PPEB 
sampling~\cite{aslamEtal08}---and analyzing the results using the Kaplan-Markov bound~\cite{stark09b} 
can lead to reasonably efficient and economical control of the familywise error rate: 
the risk that one or more incorrect election outcomes will escape a full hand count.
Compared with auditing races independently to control the risk to the same level,
the MARROP approach can reduce the expected number of batches, ballots and 
votes that need to be audited.

\bibliography{../Bib/pbsBib}
\bibliographystyle{plain}

\end{document}